\documentstyle{article}
\input tcilatex

\QQQ{Language}{
American English
}

\begin{document}

\author{Gunn A. Quznetsov}
\title{THE WEAK ISOSPIN SPACE}
\maketitle

\begin{abstract}
The Clifford pentad of the 4X4 matrices defines the 5-dimensional space.
Each weak isospin transformation divides an electron on two components,
which scatter in the 2-dimensional subspace and which indiscernible in the
orthogonal 3-dimensional subspace.

\end{abstract}

In the weak isospin theory we have got the following entities:

the right electron state vector $e_R$,

the left electron state vector $e_L$,

the electron state vector $e$ ($e=\left[ 
\begin{array}{c}
e_R \\ 
e_L 
\end{array}
\right] $),

the left neutrino state vector $\nu _L$,

the zero vector right neutrino $\nu _R$.

the unitary $2\times 2$ matrix $U$ of the isospin transformation.($\det
\left( U\right) =1$).

This matrix acts on the vectors of the kind:$\left[ 
\begin{array}{c}
\nu _L \\ 
e_L 
\end{array}
\right] $.

Therefore, in this theory: if

$$
U=\left[ 
\begin{array}{cc}
u_{1,1} & u_{1,2} \\ 
u_{2,1} & u_{2,2} 
\end{array}
\right] 
$$

then the matrix

$$
\left[ 
\begin{array}{cccc}
1 & 0 & 0 & 0 \\ 
0 & u_{1,1} & 0 & u_{1,2} \\ 
0 & 0 & 1 & 0 \\ 
0 & u_{2,1} & 0 & u_{2,2} 
\end{array}
\right] 
$$

operates on the vector

$$
\left[ 
\begin{array}{c}
e_R \\ 
e_L \\ 
\nu _R \\ 
\nu _L 
\end{array}
\right] \text{.} 
$$

Because $e_R$, $e_L$, $\nu _R$, $\nu _L$ are the two-component vectors then

$$
\left[ 
\begin{array}{c}
e_R \\ 
e_L \\ 
\nu _R \\ 
\nu _L 
\end{array}
\right] \text{ is }\left[ 
\begin{array}{c}
e_{R1} \\ 
e_{R2} \\ 
e_{L1} \\ 
e_{L2} \\ 
0 \\ 
0 \\ 
\nu _{L1} \\ 
\nu _{L2} 
\end{array}
\right] 
$$

and

$$
\left[ 
\begin{array}{cccc}
1 & 0 & 0 & 0 \\ 
0 & u_{1,1} & 0 & u_{1,2} \\ 
0 & 0 & 1 & 0 \\ 
0 & u_{2,1} & 0 & u_{2,2} 
\end{array}
\right] \text{ is }\underline{U}=\text{ }\left[ 
\begin{array}{cccccccc}
1 & 0 & 0 & 0 & 0 & 0 & 0 & 0 \\ 
0 & 1 & 0 & 0 & 0 & 0 & 0 & 0 \\ 
0 & 0 & u_{1,1} & 0 & 0 & 0 & u_{1,2} & 0 \\ 
0 & 0 & 0 & u_{1,1} & 0 & 0 & 0 & u_{1,2} \\ 
0 & 0 & 0 & 0 & 1 & 0 & 0 & 0 \\ 
0 & 0 & 0 & 0 & 0 & 1 & 0 & 0 \\ 
0 & 0 & u_{2,1} & 0 & 0 & 0 & u_{2,2} & 0 \\ 
0 & 0 & 0 & u_{2,1} & 0 & 0 & 0 & u_{2,2} 
\end{array}
\right] \text{.} 
$$

This matrix has got eight orthogonal normalized eigenvectors of kind:

$$
\underline{s_1}=\left[ 
\begin{array}{c}
1 \\ 
0 \\ 
0 \\ 
0 \\ 
0 \\ 
0 \\ 
0 \\ 
0 
\end{array}
\right] ,\underline{s_2}=\left[ 
\begin{array}{c}
0 \\ 
1 \\ 
0 \\ 
0 \\ 
0 \\ 
0 \\ 
0 \\ 
0 
\end{array}
\right] ,\underline{s_3}=\left[ 
\begin{array}{c}
0 \\ 
0 \\ 
\varpi \\ 
0 \\ 
0 \\ 
0 \\ 
\chi \\ 
0 
\end{array}
\right] ,\underline{s_4}=\left[ 
\begin{array}{c}
0 \\ 
0 \\ 
0 \\ 
\chi ^{*} \\ 
0 \\ 
0 \\ 
0 \\ 
-\varpi ^{*} 
\end{array}
\right] , 
$$

$$
\underline{s_5}=\left[ 
\begin{array}{c}
0 \\ 
0 \\ 
0 \\ 
0 \\ 
1 \\ 
0 \\ 
0 \\ 
0 
\end{array}
\right] ,\underline{s_6}=\left[ 
\begin{array}{c}
0 \\ 
0 \\ 
0 \\ 
0 \\ 
0 \\ 
1 \\ 
0 \\ 
0 
\end{array}
\right] ,\underline{s_7}=\left[ 
\begin{array}{c}
0 \\ 
0 \\ 
\chi ^{*} \\ 
0 \\ 
0 \\ 
0 \\ 
-\varpi ^{*} \\ 
0 
\end{array}
\right] ,\underline{s_8}=\left[ 
\begin{array}{c}
0 \\ 
0 \\ 
0 \\ 
\varpi \\ 
0 \\ 
0 \\ 
0 \\ 
\chi 
\end{array}
\right] . 
$$

The corresponding eigenvalues are: $1$, $1$, $\exp \left( i\cdot \lambda
\right) $, $\exp \left( i\cdot \lambda \right) $, $1$,$1$, $\exp \left(
-i\cdot \lambda \right) $, $\exp \left( -i\cdot \lambda \right) $.

These vectors constitute the orthogonal basis in this 8-dimensional space.

Let

$$
\underline{\gamma ^0}=\left[ 
\begin{array}{cc}
\gamma ^0 & O \\ 
O & \gamma ^0 
\end{array}
\right] 
$$

if $O$ is zero $4\times 4$ matrix and

$$
\gamma ^0=\left[ 
\begin{array}{cccc}
0 & 0 & 1 & 0 \\ 
0 & 0 & 0 & 1 \\ 
1 & 0 & 0 & 0 \\ 
0 & 1 & 0 & 0 
\end{array}
\right] , 
$$

and \underline{$\beta ^4$}$=\left[ 
\begin{array}{cc}
\beta ^4 & O \\ 
O & \beta ^4 
\end{array}
\right] $ if

$$
\beta ^4=\left[ 
\begin{array}{cccc}
0 & 0 & i & 0 \\ 
0 & 0 & 0 & i \\ 
-i & 0 & 0 & 0 \\ 
0 & -i & 0 & 0 
\end{array}
\right] \text{.} 
$$

The vectors $\left[ 
\begin{array}{c}
e_{R1} \\ 
e_{R2} \\ 
e_{L1} \\ 
e_{L2} \\ 
0 \\ 
0 \\ 
0 \\ 
0 
\end{array}
\right] $, $\left[ 
\begin{array}{c}
e_{R1} \\ 
e_{R2} \\ 
0 \\ 
0 \\ 
0 \\ 
0 \\ 
0 \\ 
0 
\end{array}
\right] $, $\left[ 
\begin{array}{c}
0 \\ 
0 \\ 
e_{L1} \\ 
e_{L2} \\ 
0 \\ 
0 \\ 
0 \\ 
0 
\end{array}
\right] $ correspond to the state vectors $e$, $e_R$ and $e_L$ resp.

In this case \cite{l1} \underline{$e$}$^{\dagger }\cdot \underline{\gamma ^0}%
\cdot \underline{e}=J_0$, \underline{$e$}$^{\dagger }\cdot \underline{\beta
^4}\cdot \underline{e}=J_4$, $J_0=\underline{e}^{\dagger }\cdot \underline{e}%
\cdot V_0$, $J_4=\underline{e}^{\dagger }\cdot \underline{e}\cdot V_4$.

For the vector $e$ the numbers $k_3$, $k_4$, $k_7$, $k_8$ exist, for which: 
\underline{$e$}$=(e_{R1}\cdot \underline{s_1}+e_{R2}\cdot \underline{s_2}%
)+(k_3\cdot \underline{s_3}+k_4\cdot \underline{s_4})+(k_7\cdot \underline{%
s_7}+k_8\cdot \underline{s_8})$.

Here \underline{$e_R$}$=(e_{R1}\cdot \underline{s_1}+e_{R2}\cdot \underline{%
s_2})$. If \underline{$e_{La}$}$=(k_3\cdot \underline{s_3}+k_4\cdot 
\underline{s_4})$ and \underline{$e_{Lb}$}$=(k_7\cdot \underline{s_7}%
+k_8\cdot \underline{s_8})$ then \underline{$U$}$\cdot \underline{e_{La}}%
=\exp \left( i\cdot \lambda \right) \cdot $\underline{$e_{La}$} and 
\underline{$U$}$\cdot \underline{e_{Lb}}=\exp \left( -i\cdot \lambda \right)
\cdot $\underline{$e_{Lb}$}.

Let for all $k$ ($1\leq k\leq 8$): \underline{$h_k$}$=\underline{\gamma ^0}%
\cdot $\underline{$s_k$}. The vectors \underline{$h_k$} constitute the
orthogonal basis, too. And the numbers $q_3$, $q_4$, $q_7$, $q_8$ exist, for
which: \underline{$e_R$}$=(q_3\cdot \underline{h_3}+q_4\cdot \underline{h_4}%
)+(q_7\cdot \underline{h_7}+q_8\cdot \underline{h_8})$.

Let \underline{$e_{Ra}$}$=(q_3\cdot \underline{h_3}+q_4\cdot \underline{h_4}%
) $, \underline{$e_{Rb}$}$=(q_7\cdot \underline{h_7}+q_8\cdot \underline{h_8}%
)$, \underline{$e_a$}$=\underline{e_{Ra}}+$\underline{$e_{La}$} and 
\underline{$e_b$}$=\underline{e_{Rb}}+$\underline{$e_{Lb}$}.

Let \underline{$e_a$}$^{\dagger }\cdot \underline{\gamma ^0}\cdot \underline{%
e_a}=J_{0a}$, \underline{$e_a$}$^{\dagger }\cdot \underline{\beta ^4}\cdot 
\underline{e_a}=J_{4a}$, $J_{0a}=\underline{e_a}^{\dagger }\cdot \underline{%
e_a}\cdot V_{0a}$, $J_{4a}=\underline{e_a}^{\dagger }\cdot \underline{e_a}%
\cdot V_{4a}$,

\underline{$e_b$}$^{\dagger }\cdot \underline{\gamma ^0}\cdot \underline{e_b}%
=J_{0b}$, \underline{$e_b$}$^{\dagger }\cdot \underline{\beta ^4}\cdot 
\underline{e_b}=J_{4b}$, $J_{0b}=\underline{e_b}^{\dagger }\cdot \underline{%
e_b}\cdot V_{0b}$, $J_{4b}=\underline{e_b}^{\dagger }\cdot \underline{e_b}%
\cdot V_{4b}$.

In this case: $J_0=J_{0a}+J_{0b}$, $J_4=J_{4a}+J_{4b}$.

Let $\left( \underline{U}\cdot \underline{e_a}\right) ^{\dagger }\cdot 
\underline{\gamma ^0}\cdot \left( \underline{U}\cdot \underline{e_a}\right)
=J_{0a}^{\prime }$, $\left( \underline{U}\cdot \underline{e_a}\right)
^{\dagger }\cdot \underline{\beta ^4}\cdot \underline{\left( \underline{U}%
\cdot \underline{e_a}\right) }=J_{4a}^{\prime }$, $J_{0a}^{\prime }=\left( 
\underline{U}\cdot \underline{e_a}\right) ^{\dagger }\cdot \left( \underline{%
U}\cdot \underline{e_a}\right) \cdot V_{0a}^{\prime }$, $J_{4a}^{\prime
}=\left( \underline{U}\cdot \underline{e_a}\right) ^{\dagger }\cdot \left( 
\underline{U}\cdot \underline{e_a}\right) \cdot V_{4a}^{\prime }$,

$\left( \underline{U}\cdot \underline{e_b}\right) ^{\dagger }\cdot 
\underline{\gamma ^0}\cdot \left( \underline{U}\cdot \underline{e_b}\right)
=J_{0b}^{\prime }$, $\left( \underline{U}\cdot \underline{e_b}\right)
^{\dagger }\cdot \underline{\beta ^4}\cdot \left( \underline{U}\cdot 
\underline{e_b}\right) =J_{4b}^{\prime }$, $J_{0b}^{\prime }=\left( 
\underline{U}\cdot \underline{e_b}\right) ^{\dagger }\cdot \left( \underline{%
U}\cdot \underline{e_b}\right) \cdot V_{0b}^{\prime }$, $J_{4b}^{\prime
}=\left( \underline{U}\cdot \underline{e_b}\right) ^{\dagger }\cdot \left( 
\underline{U}\cdot \underline{e_b}\right) \cdot V_{4b}^{\prime }$.

In this case:

$$
V_{0a}^{\prime }=V_{0a}\cdot \cos \left( \lambda \right) +V_{4a}\cdot \sin
\left( \lambda \right) , 
$$

$$
V_{4a}^{\prime }=V_{4a}\cdot \cos \left( \lambda \right) -V_{0a}\cdot \sin
\left( \lambda \right) ; 
$$

$$
V_{0b}^{\prime }=V_{0b}\cdot \cos \left( \lambda \right) -V_{4b}\cdot \sin
\left( \lambda \right) , 
$$

$$
V_{4b}^{\prime }=V_{4b}\cdot \cos \left( \lambda \right) +V_{0b}\cdot \sin
\left( \lambda \right) \text{.} 
$$

Hence, every isospin transformation divides a electron on two components,
which scatter on the angle $2\cdot \lambda $ in the space of ($J_0$, $J_4$).
These components are indiscernible in the space of ($j_1$, $j_2$, $j_3$) 
\cite{l2}.

My other addresses are:

gunn.q@usa.net

quznets@geocities.com

\end{document}